\documentclass[aps,plb,twocolumn,superscriptaddress,notitlepage,nofootinbib]{revtex4-1}

\usepackage{amsmath,bm}
\usepackage{graphicx}
\usepackage{epstopdf}
\usepackage{subfigure}
\usepackage{epsfig}
\usepackage{amsmath,amssymb,amsfonts}
\usepackage{color}
\usepackage[utf8]{inputenc}
\usepackage{verbatim}
\usepackage{array}
\usepackage{hyperref}
\usepackage{ulem}
\usepackage{multirow}

\usepackage{titlesec}

\begin{document}

\renewcommand\thesection{\arabic{section} }

\renewcommand{\topfraction}{0.85}
\renewcommand{\textfraction}{0.1}
\renewcommand{\floatpagefraction}{0.75}
\newcommand{\bea}{\begin{eqnarray}}
\newcommand{\eea}{\end{eqnarray}}
\newcommand{\nn}{\nonumber}

\title{Unraveling Gluon Jet Quenching through $J/\psi$  Production in Heavy-Ion Collisions}

\date{\today  \hspace{1ex}}

\author{Shan-Liang Zhang}
\affiliation{Key Laboratory of Atomic and Subatomic Structure and Quantum Control (MOE), Institute of Quantum Matter, South China Normal University, Guangzhou 510006, China}
\affiliation{Guangdong Provincial Key Laboratory of Nuclear Science, Institute of Quantum Matter, South China Normal University, Guangzhou 510006, China}
\affiliation{Guangdong-Hong Kong Joint Laboratory of Quantum Matter, Southern Nuclear Science Computing Center, South China Normal University, Guangzhou 510006, China}

\author {Jinfeng Liao }
\email[Corresponding author: ]{liaoji@indiana.edu}
\address{Physics Department and Center for Exploration of Energy and Matter, Indiana University, 2401 N Milo B. Sampson Lane, Bloomington, Indiana 47408, USA}

\author{Guang-You Qin}
\email[Corresponding author: ]{guangyou.qin@ccnu.edu.cn}
\affiliation{Key Laboratory of Quark \& Lepton Physics (MOE) and Institute of Particle Physics,
 Central China Normal University, Wuhan 430079, China}

\author{Enke Wang }
\email[Corresponding author: ]{wangek@scnu.edu.cn}
\affiliation{Key Laboratory of Atomic and Subatomic Structure and Quantum Control (MOE), Institute of Quantum Matter, South China Normal University, Guangzhou 510006, China}
\affiliation{Guangdong Provincial Key Laboratory of Nuclear Science, Institute of Quantum Matter, South China Normal University, Guangzhou 510006, China.}
\affiliation{Guangdong-Hong Kong Joint Laboratory of Quantum Matter, Southern Nuclear Science Computing Center, South China Normal University, Guangzhou 510006, China}

\author{Hongxi Xing}
\email[Corresponding author: ]{hxing@m.scnu.edu.cn}
\affiliation{Key Laboratory of Atomic and Subatomic Structure and Quantum Control (MOE), Institute of Quantum Matter, South China Normal University, Guangzhou 510006, China}
\affiliation{Guangdong Provincial Key Laboratory of Nuclear Science, Institute of Quantum Matter, South China Normal University, Guangzhou 510006, China}
\affiliation{Guangdong-Hong Kong Joint Laboratory of Quantum Matter, Southern Nuclear Science Computing Center, South China Normal University, Guangzhou 510006, China}
\affiliation{Southern Center for Nuclear-Science Theory (SCNT), Institute of Modern Physics, Chinese Academy of Sciences, Huizhou 516000, China}

\begin{abstract}
Jet quenching has long been regarded as one of the key signatures for the formation of quark-gluon plasma  in heavy-ion collisions. Despite significant efforts, the separate identification of quark and gluon jet quenching has remained as a challenge. Here we show that $J/\psi$ in high transverse momentum ($p_\text{T}$) region provides a uniquely sensitive probe of in-medium gluon energy loss since its production at high $p_\text{T}$ is particularly dominated by gluon fragmentation. Such gluon-dominance is first demonstrated for  the baseline of proton-proton collisions within the framework of leading power NRQCD factorization formalism. We then use the linear Boltzmann transport model combined with hydrodynamics  for the simulation of jet-medium interaction in nucleus-nucleus collisions. The satisfactory description of experimental data on both nuclear modification factor $R_{\text{AA}}$ and elliptic flow $v_2$ reveals, for the first time, that the  gluon jet quenching is the driving force for high $p_\text{T}$ $J/\psi$ suppression. This novel finding is further confirmed by the data-driven Bayesian analyses of relevant experimental measurements, from which we also obtain the first quantitative extraction of the gluon energy loss distribution in the quark-gluon plasma.\\


\end{abstract}


\maketitle

\section{Introduction}

Since the start of the pioneering high energy nuclear collision experiments at Relativistic Heavy Ion Collider (RHIC), followed by the cutting edge measurements at the Large Hadron Collider (LHC), many experimental signatures have been suggested for the discovery of Quark-Gluon Plasma (QGP), a new form of matter under extreme high temperature. Among these signatures, jet quenching has been widely considered as one of the most important probes~\cite{Qin:2015srf}, which results in the yield suppression of high $p_\text{T}$ hadrons and jets~\cite{Qin:2007rn,Majumder:2011uk,ALICE:2010yje}, the shift of two-particle correlations~\cite{PHENIX:2009cvn,Zhang:2007ja}, the modification of jet internal structure~\cite{ATLAS:2019dsv,Chang:2019sae}, as well as azimuthal anisotropy ($v_2$) of hadrons and jets~\cite{STAR:2003wqp,CMS:2017xgk} in the large transverse momentum ($p_\text{T}$) region in nucleus-nucleus (AA) collisions, in comparison with an equivalent number of proton-proton (pp) collisions.

While comprehensive efforts have been devoted to extract key transport properties of QGP based on jet quenching frameworks (see e.g. by JET and JETSCAPE collaborations~\cite{JET:2013cls,JETSCAPE:2020shq}), the information on the jet quenching of specific parton type is still limited. A separate determination of quark and gluon jet energy loss could play a significant role in revealing the fundamental color structures of the QGP and testing the color representation dependence of the jet-medium interaction~\cite{Frye:2017yrw,Gras:2017jty}.
In particular, such knowledge  can help directly access and verify the unique non-Abelian gauge symmetry of quantum chromodynamics (QCD) which is a key component of the Standard Model as fundamental laws of nature.
 This however proves difficult, as quark and gluon contributions are often entangled together in final state hadronic observables. A clean method for identifying quark or gluon energy loss remains a challenge, despite many past attempts such as the multivariate analysis of jet substructure observables~\cite{Chien:2018dfn}, the proposal of using the averaged jet charge~\cite{Chen:2019gqo,CMS:2020plq} and electroweak boson tagged jet~\cite{Zhang:2018urd,CMS:2017eqd}, as well as data driven analysis of Casimir factors and parametrizations of jet quenching models~\cite{Apolinario:2020nyw,Arleo:2017ntr,Spousta:2016agr}.

In this Letter, we demonstrate that $J/\psi$ production at high $p_\text{T}$ can serve as a unique probe of gluon in-medium energy loss. It is well known that heavy quarkonia production at low $p_\text{T}$ have long been studied as a signature of  color deconfinement~\cite{Matsui:1986dk} and a thermometer of QGP. Experimental data for low-$p_\text{T}$ $J/\psi$ suppression in Pb+Pb~\cite{ALICE:2012jsl} and Au+Au collisions~\cite{PHENIX:2006gsi} are well explained by the interplay between the dissociation and regeneration of quarkonium states heavy quark pairs~\cite{Braun-Munzinger:2000csl,Thews:2000rj} in hot QGP medium.
On the other hand, $J/\psi$ suppression at large $p_\text{T}$~\cite{CMS:2016mah,ATLAS:2018hqe} was described by the transport models~\cite{Zhou:2014kka,He:2021zej} and effective theory approaches~\cite{Aronson:2017ymv,Makris:2019ttx}. However, which physics effect ``dictates'' the suppression of $J/\psi$ in the standard jet quenching picture at high $p_\text{T}$? This question motivates the present work.

To benchmark the nuclear effect at high $p_\text{T}$, we first validate the pp baseline for $J/\psi$ production by utilizing a theoretical framework analogous to the gluon fragmentation improved PYTHIA (GFIP) method~\cite{Bain:2017wvk},
in which the heavy quarkonia are produced by the fragmentation of shower quarks and gluons. The jet evolution in QGP medium is simulated by the linear Boltzmann transport (LBT) model~\cite{ He:2015pra,Cao:2016gvr}, which includes both elastic and inelastic interactions between the fast jet partons and the medium constituents. We demonstrate, for the first time, that the suppression of high $p_\text{T}$ $J/\psi$ yield in heavy-ion collisions is dominated by the gluon jet quenching effect. Therefore, measurements of high $p_\text{T}$ $J/\psi$ production can be used as  a unique probe to identify gluon energy loss. Such a novel finding is further verified by the data-driven Bayesian analysis of the relevant experimental data, from  which we perform the first quantitative extraction of the gluon energy loss distribution in quark-gluon plasma.

\section{Theoretical frameworks}
In the high transverse momentum region ($p_\text{T}\gg m_\text{c}$ with $m_\text{c}$ the charm quark mass), one encounters large logarithms of $p_\text{T}^2/m_\text{c}^2$ in fixed order calculations using the nonrelativistic QCD (NRQCD), a widely adopted approach for studying heavy quarkonium production. Such large logarithms need to be resumed to preserve the convergence of perturbation theory. In the leading power of $p_\text{T}^2/m_\text{c}^2$, the cross section for hadroproduction of $J/\psi$ is schematically given by~\cite{Collins:1981uw}:
\bea
\label{eq-factorization}
&&d\sigma[AB\to J/\psi+X]
=\sum_{i} d \hat\sigma_{AB \to i+X}\otimes D_{i\to J/\psi},
\eea
where $d\hat\sigma_{AB\to i+X}$ is the cross section for inclusive parton $i$ in $AB$ collisions, and $D_{i\to J/\psi}$ is the fragmentation function (FF) for parton $i$ to produce a $J/\psi$~\cite{ Bodwin:1994jh}.

In Eq. (\ref{eq-factorization}), a standard method to resum the large logarithms of $p_\text{T}^2/m_\text{c}^2$ is achieved by solving the DGLAP evolution equation ~\cite{Gribov:1972ri,Dokshitzer:1977sg,Altarelli:1977zs}of $D_{i\to J/\psi}$ at scale $\mu_\text{f}\sim p_\text{T}$, and the corresponding $d\hat\sigma_{AB\to i+X}$ must be evaluated by the convolution of fixed order hard part coefficient and the parton distribution functions. Alternatively, the large logs can be effectively resummed using the so-called GFIP event generator~\cite{Bain:2017wvk}, where the parton $i$ production $d\hat\sigma_{AB\to i+X}$ is simulated in PYTHIA by considering the hard process followed by parton shower down to scale $2m_c$, and the corresponding $J/\psi$ fragmentation function $D_{i\to J/\psi}$ at $\mu\sim 2m_\text{c}$ can be further factorized as follows
\bea
\label{eq-Dfac}
D_{i\to J/\psi}(z,\mu) = \sum_n\hat d_{i\to [Q\bar Q(n)]}(z,\mu)\langle \mathcal{O}_{[Q\bar Q(n)]}^{J/\psi}\rangle,
\eea
where the quantum numbers $n$ for the intermediate nonrelativistic $Q\bar Q$ states are denoted as $n={}^{2S+1}L_J^{[1,8]}$ with the superscript $[1]$ or $[8]$ standing for color singlet or octet state respectively. The short-distance coefficients $\hat d_{i\to [Q\bar Q(n)]}$ are perturbatively calculable within NRQCD and can be found in Refs.~\cite{Ma:2013yla}. Contributions to $J/\psi$ production from quarks other than $c$ in the hard process are suppressed, due to two orders of magnitude suppression of light-quark fragmentation functions~\cite{Baumgart:2014upa}. Therefore, one can neglect their contribution. $\langle \mathcal{O}_{[Q\bar Q(n)]}^{J/\psi} \rangle$ is the nonperturbative long distance matrix element (LDME) representing the transition from $[Q\bar Q(n)]$ state to $J/\psi$. The GFIP event generator was shown to give good agreement with analytical calculations at next-to-leading-log-prime accuracy and
remedies the default PYTHIA in describing the LHCb data for $J/\psi$ production in jets~\cite{Bain:2017wvk}. 
Here we will extend the GFIP, as sketched in Eqs. (\ref{eq-factorization},\ref{eq-Dfac}), to the description of $J/\psi$ production  in high-$p_\text{T}$ region. In particular, we use MadGraph~\cite{Alwall:2014hca} for the hard parton creation and PYTHIA8 for parton shower, with the LDMEs taken from~\cite{Bodwin:2014gia}. To be consistent with Ref.~\cite{Bodwin:2014gia} and ensure the validity of Eq. (\ref{eq-factorization}), in principle, the gluon splitting and fragmentation processes must happen outside the QGP medium. 
Based on the formation time argument, the minimum $p_\text{T}$ of the gluons (and similarly $J/\Psi$) is estimated to be $p_{{\rm T,min}} \sim 2m_\text{c}^2 L \sim 50$~GeV for a typical path length $L\sim 3$~fm (considering the rapid expansion and cooling of the medium). 
It is interesting to note that the energy loss of a gluon before splitting into heavy quark pairs is not very different from the total energy loss experienced by a pair of splitted heavy quarks.  
In this paper, we restrict our discussion in the high $p_\text{T}$ region ($p_\text{T} > 10$~GeV) and focus on direct $J/\psi$ production.
While the feed-down from higher charmonium states contributes an appreciable fraction in the total prompt $J/\Psi$ production~\cite{Faccioli:2008ir}, their contribution in jet quenching observable like $R_{\text{AA}}$ would largely cancel out  in the ratio of AA and pp cross sections.

In relativistic heavy ion collisions, the energetic parton $i$ from hard interaction encounters multiple scatterings inside the QGP medium  before its fragmentation into high-$p_\text{T}$ heavy quarkonium. A Linear Boltzmann Transport (LBT) model is employed to incorporate both elastic and inelastic processes for the charm quarks and gluons scattering with medium constituents~\cite{ He:2015pra,Cao:2016gvr}. For elastic scatterings, the evolution of hard partons are simulated by the following linear Boltzmann transport equation,
\begin{eqnarray}
&p_1\cdot\partial f_a(p_1)=-\int\frac{d^3p_2}{(2\pi)^32E_2}\int\frac{d^3p_3}{(2\pi)^32E_3}\int\frac{d^3p_4}{(2\pi)^32E_4} \nonumber \\
&\frac{1}{2}\sum _{b(c,d)}[f_a(p_1)f_b(p_2)-f_c(p_3)f_d(p_4)]|M_{ab\rightarrow cd}|^2 \nonumber \\
&\times S_2(s,t,u)(2\pi)^4\delta^4(p_1+p_2-p_3-p_4) + {\rm inel.}
\end{eqnarray}
where $f_{i=a,b,c,d}$ are the phase-space distributions of jet shower partons ($a, c$) and medium partons ($b, d$), $|M_{ab\rightarrow cd}|$ are the corresponding elastic matrix elements which are regulated by a Lorentz-invariant regulation condition $S_2(s,t,u)=\theta(s>2\mu^{2}_{\text{D}})\theta(-s+\mu^{2}_{\text{D}}\leq t \leq -\mu^{2}_{\text{D}})$. $\mu_{\text{D}}^{2}=g^{2}T^{2}(N_{\text{c}}+N_{\text{f}}/2)/3$ is the Debye screening mass. The effect of  inelastic scatterings is described by the higher-twist formalism for induced gluon radiation as follows~\cite{Guo:2000nz,Zhang:2003wk},
\begin{equation}
\frac{dN_g}{dxdk_\perp^2 dt}=\frac{2\alpha_sC_AP(x)\hat{q}}{\pi k_\perp^4}\left(\frac{k_\perp^2}{k_\perp^2+x^2M^2}\right)^4\sin^2\left(\frac{t-t_i}{2\tau_f}\right)  .
\end{equation}
Here $x$ denotes the energy fraction of the radiated gluon relative to a parent parton with mass $M$, $k_\perp$ is the transverse momentum. A lower energy cut-off $x_{\text{min}}=\mu_{\text{D}}/E$ is applied for the emitted gluon in the calculation. $P(x)$ is the splitting function in vacuum, and $\tau_f=2Ex(1-x)/(k^2_\perp+x^2M^2)$ is the formation time of the radiated gluons in QGP. 
More details on the implementation of the LBT model with the HT formalism can be found in Refs.~\cite{He:2015pra, Cao:2016gvr}.
In principle, one can also use other formalisms to calculate medium-induced gluon radiation and parton energy loss, such as BDMPS-Z~\cite{Baier:1996kr, Zakharov:1996fv}, ASW~\cite{Wiedemann:2000za}, GLV~\cite{Gyulassy:2000er}, AMY~\cite{Arnold:2001ms}. For a comprehensive comparison of various energy loss formalisms, the reader is referred to Ref.~\cite{Armesto:2011ht}.
The dynamic evolution of bulk medium is given by 3+1D CLVisc hydrodynamical model~\cite{Pang:2012he} with parameters fixed by reproducing hadron spectra from experimental measurements. 
In LBT model, the strong coupling constant $\alpha_\text{s}$ is the only free parameter that controls the strength of parton-medium interaction. We follow the previous study~\cite{Zhang:2018urd} and set $\alpha_s=0.18$; There is no additional parameter in our calculation.
Thus, the jet transport parameter $\hat q$ is the same as previous LBT calculations for various jet quenching observables, such as light and heavy flavor hadrons suppression, single inclusive jets suppression, as well as boson-jet correlation~\cite{He:2015pra,Cao:2016gvr,  Zhang:2018urd}.

\section{Numerical results}
\begin{figure}
  \centering
 \includegraphics[scale=0.3]{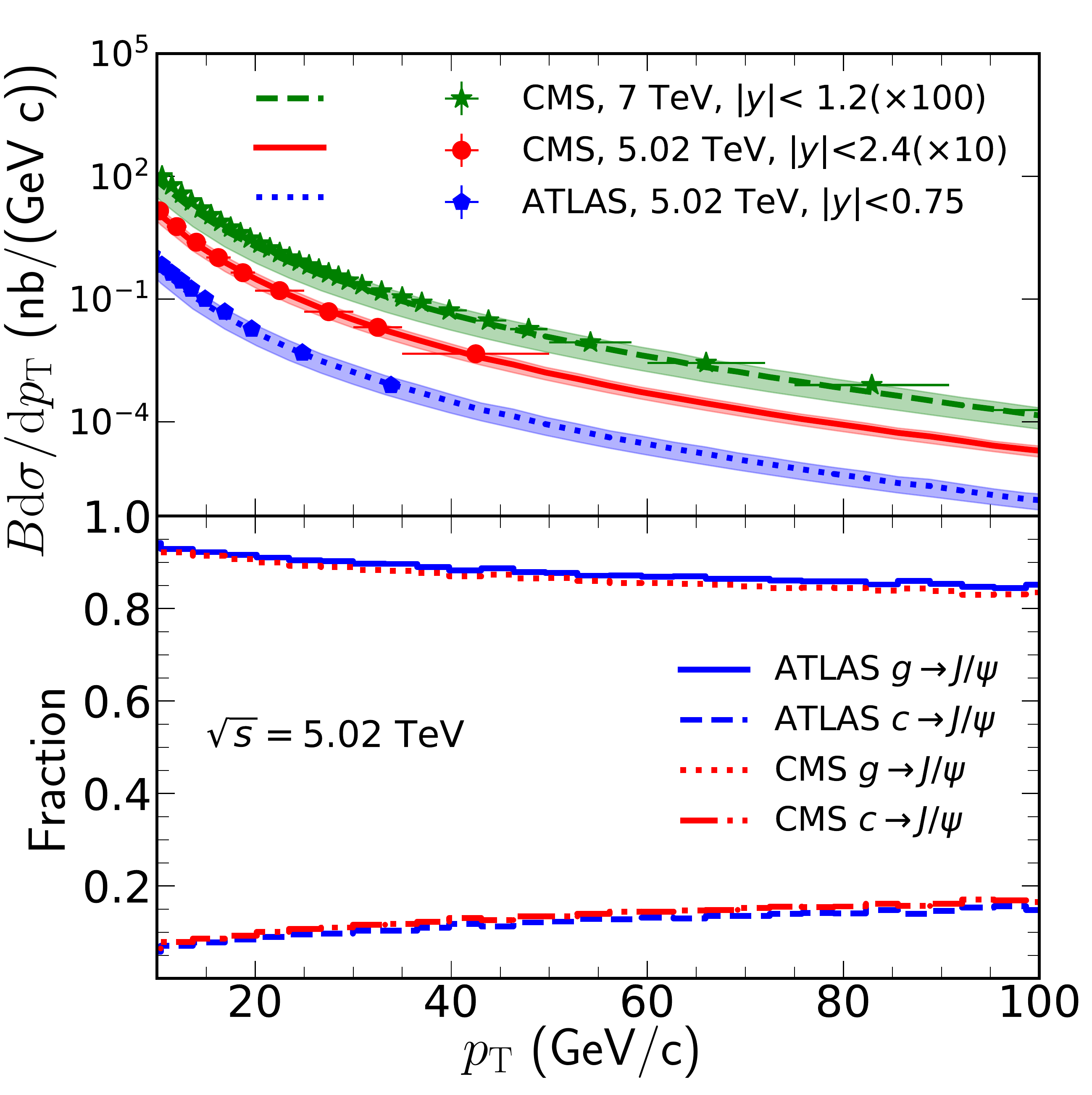}\vspace{5pt}
  \caption{(Color online) Transverse momentum spectra of $J/\psi$  production in p+p collisions  from leading power NRQCD calculations and the comparison with CMS and ATLAS data~\cite{ATLAS:2018hqe,CMS:2017uuv,CMS:2015lbl} . The band is the uncertainties
   of LDME. Contributions from different fragmentation channel are also shown.   }
   \label{fig-pp}
   \vspace{-10pt}
\end{figure}

We first examine the $J/\psi$ production mechanism in p+p collisions. In Fig. \ref{fig-pp} (upper panel), we show the transverse momentum spectra of $J/\psi$ production in p+p collisions at 5.02 TeV and 7 TeV from our GFIP simulations, which compare well with ATLAS~\cite{ATLAS:2018hqe} and CMS~\cite{CMS:2017uuv, CMS:2015lbl} measurements at large $p_\text{T}$. The lower panel shows the relative contributions, $f_{i\rightarrow J/\psi}=\sigma_{i\rightarrow J/\psi}/\sigma^{\text{total}}_{J/\psi}$, from gluon and charm-quark fragmentation, respectively. One can see that the contribution from charm fragmentation to the production of $J/\psi$ is less than $15\%$ over a broad range of $p_\text{T}$, indicating that high-$p_\text{T}$ $J/\psi$ production is dominated by gluon fragmentation~\cite{Baumgart:2014upa,Liu:2006hc}. This observation is also consistent with the fixed order NRQCD calculations showing that high-$p_\text{T}$ $J/\psi$ is dominated by color-octet fragmentation~\cite{Bodwin:2015iua}.

\begin{figure}
  \centering
 \includegraphics[scale=0.3]{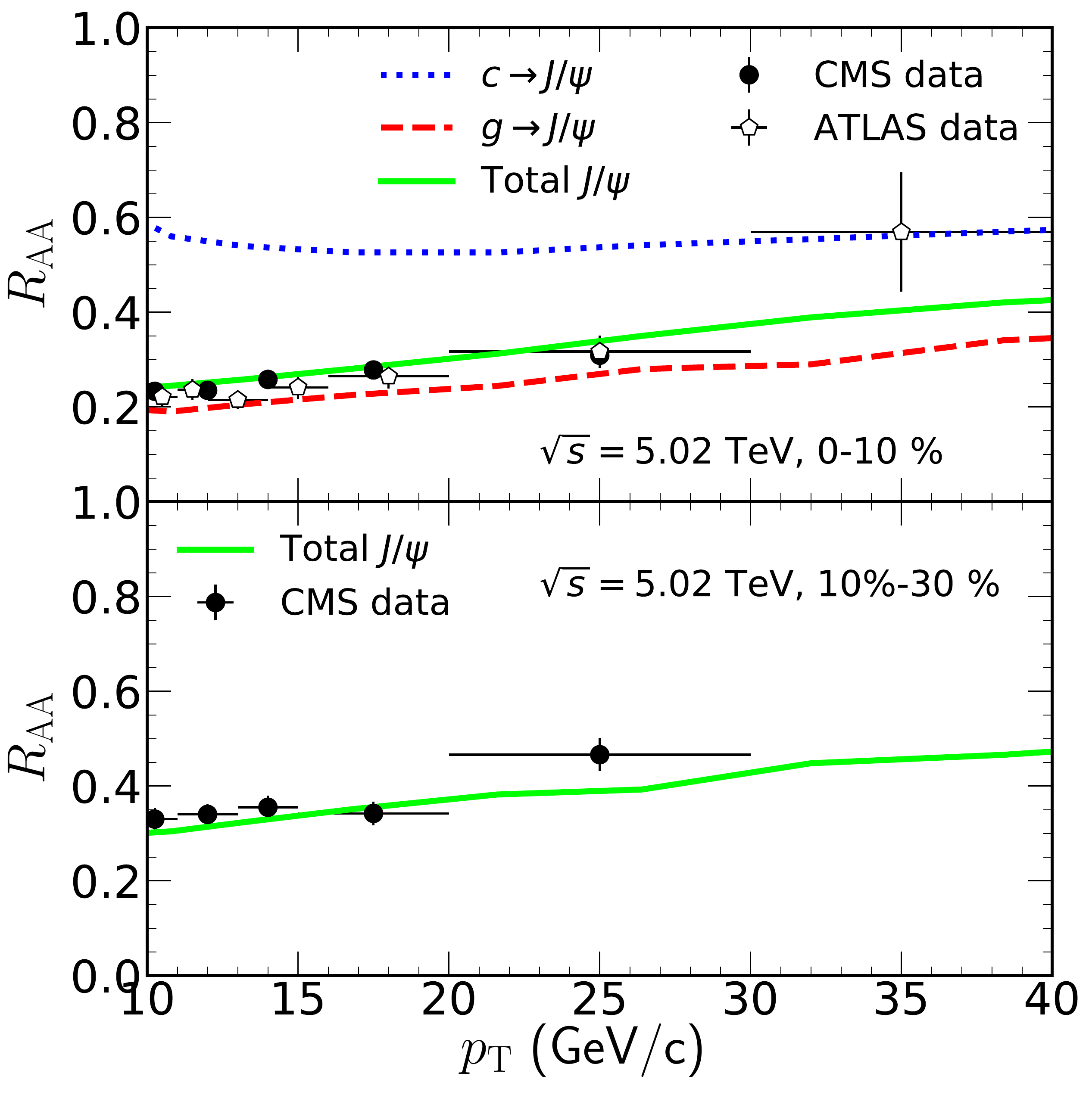}\vspace{5pt}
  \caption{(Color online)  Nuclear modification factor $R_{\text{AA}}$ evaluated as a function of  $J/\psi$ transverse momentum $p_\text{T}$ and the comparison with CMS~\cite{CMS:2017uuv} and ATLAS~\cite{ATLAS:2018hqe} experimental data in central 0-10$\%$ (top panel)  and 10$\%$-30$\%$ (bottom panel) Pb+Pb collisions at 5.02 TeV.}
  \label{fig-RAA}
  \vspace{-10pt}
\end{figure}

In Fig.~\ref{fig-RAA}, we present the nuclear modification factor $R_{\text{AA}}$ from our LBT simulations as a function of $p_\text{T}$ for $J/\psi$ production in  0-10$\%$ and 10$\%$-30$\%$ Pb+Pb collisions at 5.02 TeV, compared with CMS~\cite{CMS:2017uuv} and ATLAS~\cite{ATLAS:2018hqe} data. The total $R_{\text{AA}}$ results (shown in green lines) give a very good description of the experimental data in high-$p_\text{T}$ region for both centrality classes  within the current statistical uncertainties. We further examine the $R_{\text{AA}}$ results separately from charm quarks and gluons, in blue and red lines, respectively. A particularly interesting observation is that the strong suppression of high $p_\text{T}$ $J/\psi$ production in AA collisions is mainly driven by the gluon jet quenching effect. This result can be well explained by the dominance of gluon fragmentation in $J/\psi$ production (see Fig. \ref{fig-pp}) combined with the stronger energy loss for gluon jet than charm jet due to their different color charges. Such a novel finding is different from the naive expectation that $J/\psi$ suppression is driven by its constituent charm quarks. Our result presents a unique opportunity to directly access the gluon jet quenching by utilizing existing and future heavy ion  measurements on $J/\psi$ production at high $p_\text{T}$ region.

\begin{figure}
  \centering
  \vspace{10pt}
  \includegraphics[scale=0.26]{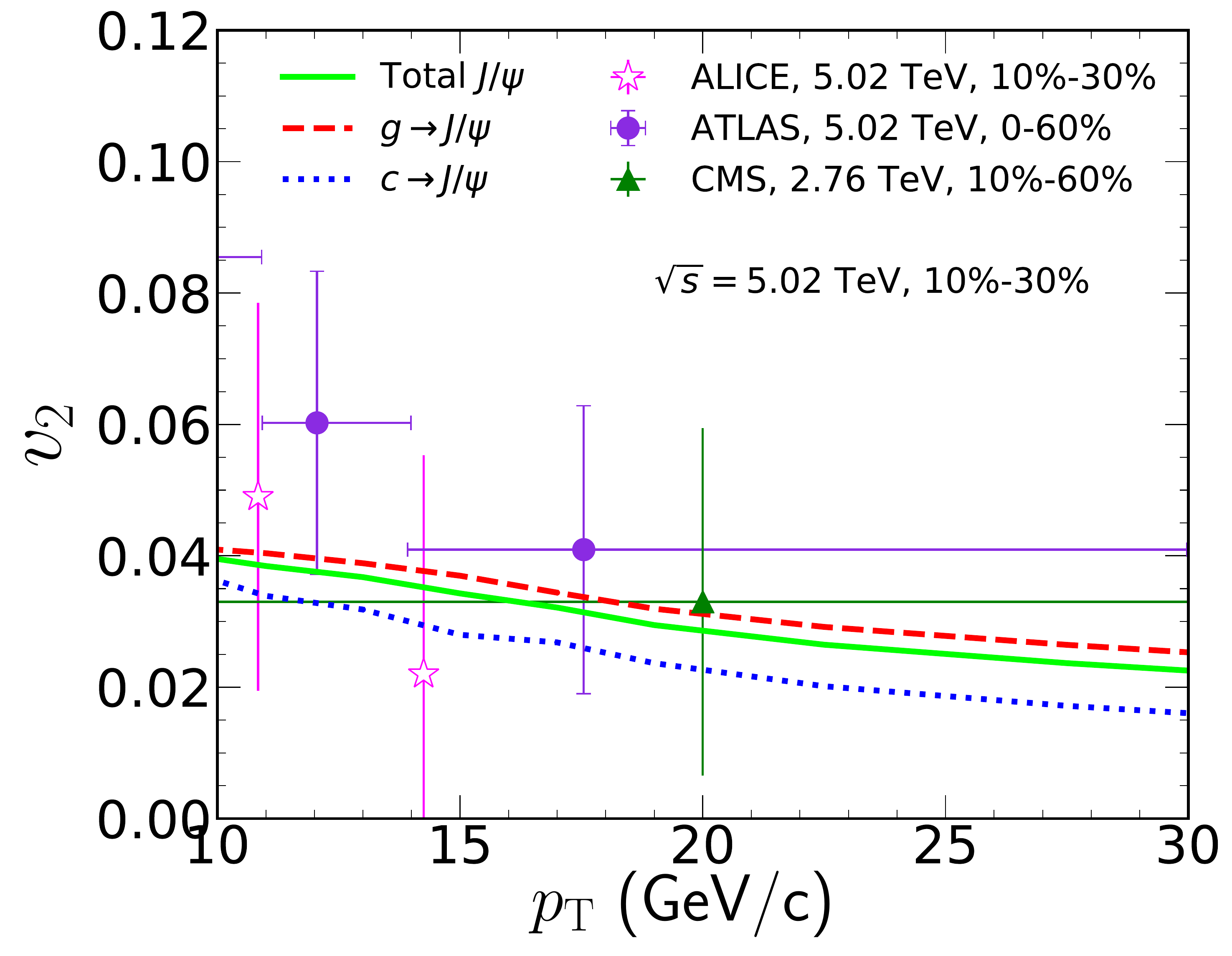}   \vspace{-5pt}
   \caption{(Color online) Elliptic flow coefficient $v_2$ of $J/\psi$ from the fragmentation of gluon (Short-dashed red line) and charm (Short-dotted dark line), and total $J/\psi$ (Solid blue line) as well as the comparison to experimental measurements~\cite{CMS:2016mah,ATLAS:2018xms,ALICE:2020pvw}.
   }
   \label{fig-v2}
  \vspace{-0pt}
\end{figure}

Another key observable for hard probes is the elliptic anisotropy coefficient $v_2$, which can be evaluated as
\begin{eqnarray}
& v_2(p_\text{T}) =  \frac{1}{d\sigma^{J/\psi}(p_\text{T}) } \sum_{i}\int d \sigma^i (\frac{p_\text{T}}{z}) v_2^i(\frac{p_\text{T}}{z})  \otimes D_{i\rightarrow J/\psi},
\label{eq-v2}
\end{eqnarray}
where $v_2^{i}$ is the elliptic flow coefficients for the parent charm quarks and gluons. As shown in Eq. (\ref{eq-v2}), the elliptic flow coefficient $v_2$ of $J/\psi$ should inherit its parent partons that fragment into it. The numerical results for $v_2$ are shown in Fig.~\ref{fig-v2} and compared to experimental data~\cite{CMS:2016mah,ATLAS:2018xms,ALICE:2020pvw}. Our results show reasonable agreement with  experimental data at high $p_\text{T}$ region which though have large uncertainties. One can see again that the contribution from gluon quenching effect dominates the $v_2$ of high-$p_\text{T}$ $J/\psi$.
\section{Bayesian extraction of gluon energy loss distribution}
To test the robustness of  the above finding based on the framework of leading power NRQCD and LBT model, we now use an alternative approach, i.e. the data-driven Bayesian analysis~\cite{Andrieu}, to examine the sensitivity of high-$p_\text{T}$ $J/\psi$ suppression to parton flavors. 
The Bayesian analyses have been successfully employed to extract the  equation of state of QCD matter~\cite{Pratt:2015zsa}, heavy quark transport coefficients~\cite{Xu:2017obm}, and  jet energy loss distributions~\cite{He:2018gks} as well as jet transport coefficient $\hat{q}$ ~\cite{JETSCAPE:2021ehl} in heavy-ion collisions.
 Here we also adopt such an advanced statistical tool  to constrain the flavor dependence of energy loss distribution from relevant experiment data.

To do that, we extend the method in Ref.~\cite{He:2018gks} to the heavy quarkonium problem and consider systematically the flavor (charm quark and gluon) dependence of parton energy loss. The production spectrum can be expressed as the convolution of its cross section in pp collisions and a flavor-dependent parton energy loss distribution
\begin{eqnarray}
&\frac{d \sigma_{\text{AA}}}{ dp_\text{T}}=  \sum_i \int \frac{d\Delta p_\text{T}^i}{\langle \Delta p_\text{T}^i\rangle} \frac{d \sigma_{\text{pp}}^i(p_\text{T}+\Delta p_\text{T}^i)}{dp_\text{T}} W^i(x)\otimes D_{i\rightarrow J/\psi},
\label{eq-sigma_AQ}
\end{eqnarray}
where $d\sigma_{\text{pp}}^i$  is the  differential cross section for parton $i$, $\Delta p_\text{T}^i$ is the energy loss of parton $i$ with initial transverse energy $p_\text{T}+\Delta p_\text{T}^i$, $ x=\Delta p_\text{T}^i/\langle \Delta p_\text{T}^i\rangle$ is the scaled energy loss, and the averaged parton energy loss is parametrized as $\langle \Delta p_\text{T}^i\rangle ( p_\text{T}^i)=\beta_i ({p_\text{T}^i}/{p_\text{T}^{0}})^{\gamma_{i}} \log({p_\text{T}^i}/{p_\text{T}^{0}})$ with $p_\text{T}^{0}=1$ GeV. 
One can see that parton energy loss directly controlls the quenching of high $p_\text{T}$ hadrons; it is also closely related to jet transport coefficient $\hat{q}$ (see e.g., Ref.~\cite{Baier:1996sk}).
In Eq. (\ref{eq-sigma_AQ}), we assume a general functional form for the scaled energy loss distribution of parton $i$~\cite{He:2018gks}:
\begin{equation}\label{straight}
W^i(x)=\frac{\alpha_{i}^{\alpha_{i}} x^{\alpha_i-1}e^{-\alpha_i x} }{\Gamma(\alpha_i)}.
\end{equation}
where $\Gamma$ is the standard Gamma-function, and the above functional form
can be empirically interpreted as the energy loss distribution resulting from $\alpha_i$ number of jet-medium scatterings~\cite{He:2018gks}. In the end, the problem is then reduced to the determination of six free parameters [$\alpha_i$, $\beta_i$, $\gamma_i$], with $i$ standing for gluon and charm quark.
It may be noted that the Bayesian analysis here uses specific functional form for the parameterzation, thus introducing long-range correlations in the parameter space which may potentially bias the extracted parameters. A possible solution to tackle such issue is to use information field based approach as presented in Ref.~\cite{Xie:2022ght}.

\begin{figure}
\vspace{5pt}
  \centering
  \includegraphics[scale=0.4]{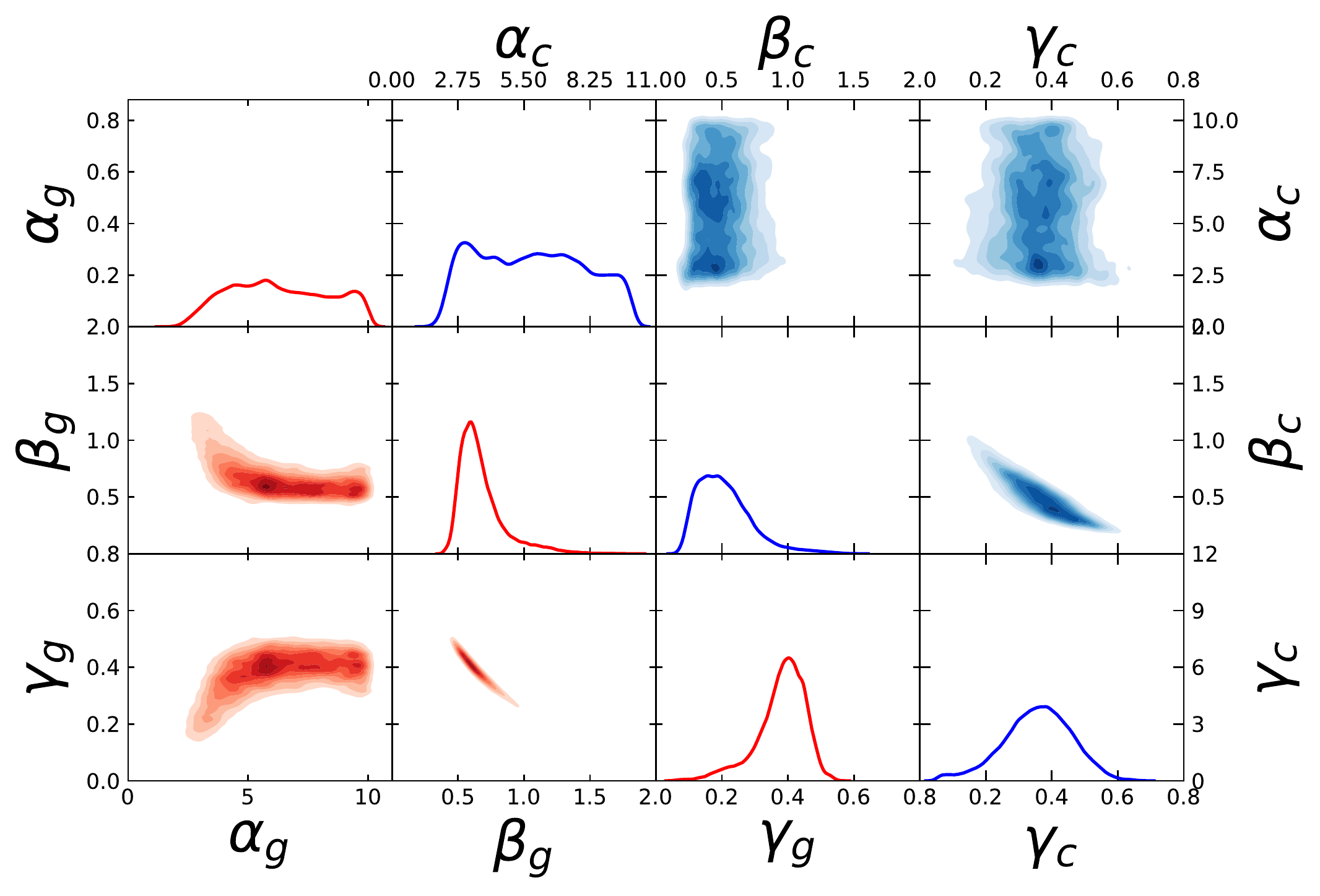}   \vspace{-5pt}
   \caption{(Color online) Density distribution (line) of and correlations (contour) between the parameters $[\alpha_i, \beta_i, \gamma_i]$ for gluon (left) and  charm quark (right) from  Bayesian  fits to experimental data on inclusive $J/\psi$ suppression in 0-$10\%$ central Pb+Pb collisions at 5.02 TeV~\cite{CMS:2017uuv, ATLAS:2018hqe}.   }\label{fig-correlation}
  \vspace{-5pt}
\end{figure}

To proceed, a uniform prior distribution $P(\theta)$ in the region $[\alpha_i, \beta_i, \gamma_i] \in [(0,10),(0,8),(0,0.8)]$ is used for the Bayesian analysis. We first run $1\times 10^6$ burn-in MCMC steps to allow the chain to reach equilibrium, and then generate
$1 \times 10^6$ MCMC steps in parameter space. We show in Fig. \ref{fig-correlation} the density distribution and correlations of the parameters [$\alpha_i$, $\beta_i$, $\gamma_i$] for gluon and charm quark from Bayesian fits to experimental data on inclusive $J/\psi$ suppression in 0-10$\%$ central Pb+Pb collisions at 5.02 TeV. One can see that the parameters for gluon energy loss are much better constrained than that for charm quark. This clearly confirms our earlier finding that $J/\psi$ production is particularly sensitive to the gluon energy loss. The analysis also shows a strong inverse correlation between $\beta_i$ and $\gamma_i$ parameters for both gluon and charm quark, in consistency with the pattern for flavor-averaged jet energy loss seen in Ref.~\cite{He:2018gks}.

\begin{figure}
  \centering
  \includegraphics[scale=0.25]{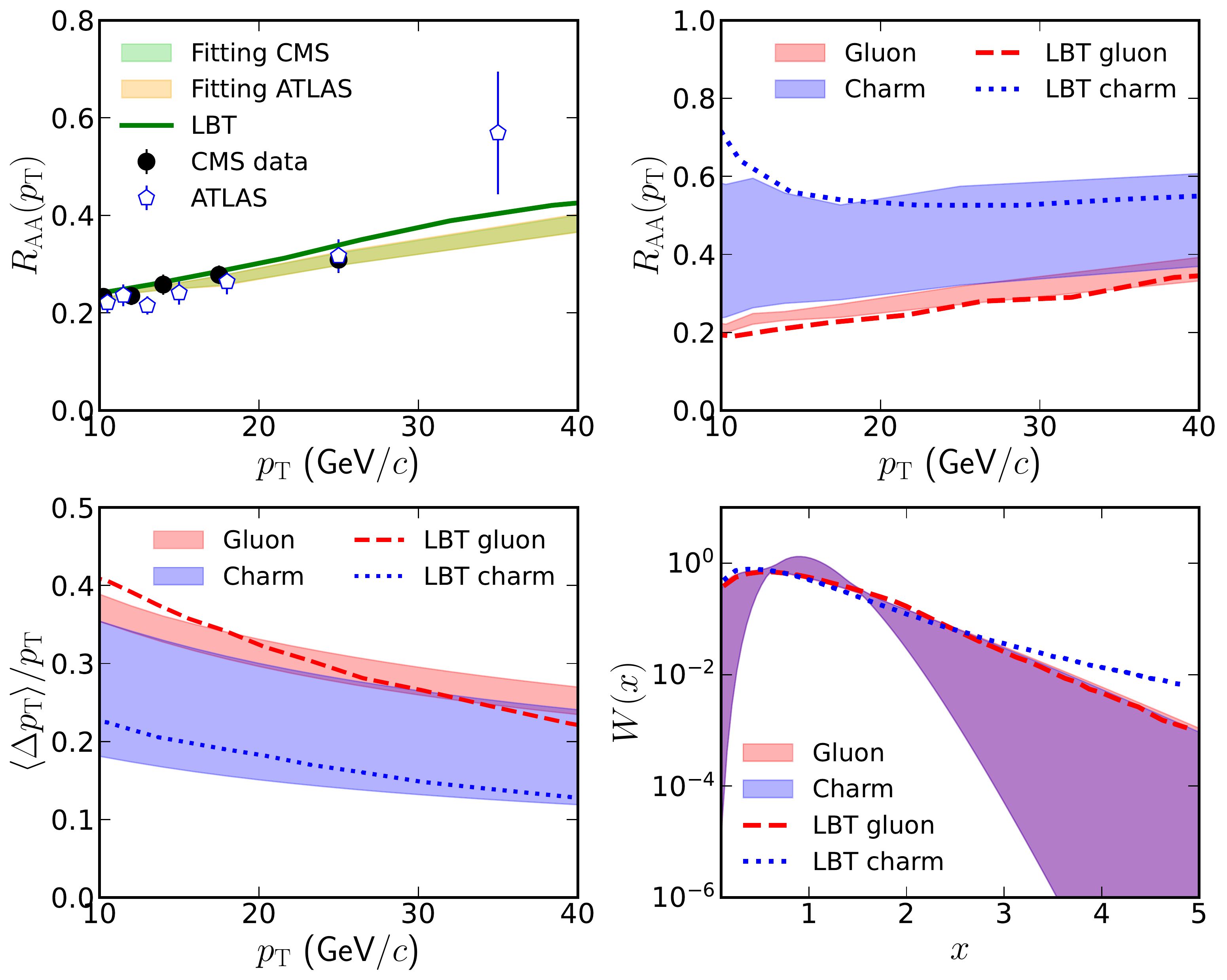}   \vspace{-5pt}
   \caption{(Color online) Nuclear modification factor for inclusive $J/\psi$ suppression, the extracted $R_{\text{AA}}$ of gluon and quarks, the extracted  fraction of average energy loss  $\langle \Delta p_\text{T}\rangle/ p_\text{T}$ and energy loss distributions $W(x)$  from Bayesian  fits to experimental data in 0-10$\%$ central Pb+Pb collisions at 5.02 TeV from CMS~\cite{CMS:2017uuv} and ATLAS~\cite{ATLAS:2018hqe}. The  bands are results of  Bayesian analysis  with one sigma deviation from the average fits of inclusive $J/\psi$ $R_{\text{AA}}$ data.
   The solid, dotted and dashed lines are from LBT simulations. }\label{fit_raa}
  \vspace{-10pt}
\end{figure}

The final results from Bayesian analysis of CMS~\cite{CMS:2017uuv} and ATLAS~\cite{ATLAS:2018hqe} data for the inclusive $J/\psi$ $R_{\text{AA}}$ as well as the extracted $R_{\text{AA}}$ separately from gluon and charm quark are shown in Fig.~\ref{fit_raa} (upper panels). The extracted average fractional energy loss $\langle \Delta p_\text{T}\rangle/ p_\text{T}$ as a function of parton energy $p_\text{T}$ and the energy loss distributions $W(x)$ are shown in Fig.~\ref{fit_raa} (lower panels), with the obtained parameters for gluon and charm quark energy loss distribution summarized in Table \ref{table:parameters}. 
Fig.~\ref{fit_raa} also shows the LBT results which roughly agree with the Bayesian results.
Our flavor-dependent results clearly demonstrate that the high-$p_\text{T}$ $J/\psi$ suppression is mainly sensitive to gluon energy loss. 
Owing to this sensitivity, the extracted average fractional energy loss $\langle \Delta p_\text{T}\rangle/ p_\text{T}$ for gluons are much better constrained than the charm quark. 
As for the energy loss distributions $W(x)$, the uncertainties for gluons and charm quarks are both quite large. We have checked that by reducing the uncertainties of experimental data by a factor of two, the gluon energy loss distribution is much better constrained.
In short, the Bayesian analysis provides a robust way to confirm that high $p_\text{T}$ $J/\psi$ suppression is particularly sensitive to gluon jet quenching and thus allows a quantitative extraction of the in-medium gluon energy loss. 

\begin{table}[!t]
  \caption{Parameters $[\alpha_i, \beta_i, \gamma_i]$ for gluon and charm quark energy loss  from  Bayesian analysis of experimental data on inclusive $J/\psi$ suppression~\cite{CMS:2017uuv, ATLAS:2018hqe} in 0-10$\%$ Pb+Pb collisions at $\sqrt{s_{\text{NN}}}$ = 5.02 TeV . 
  }
\begin{center}

  \begin{tabular}{p{1.5cm}<{\centering} p{2.cm}<{\centering} p{2.cm}<{\centering} p{2.cm}<{\centering}}
   \hline
            & $\alpha$          &    $\beta $        &  $\gamma$    \\ \hline
   \multirow{1}*{Gluon} & 5.25$\pm$1.09  &    0.7$\pm$0.07 & 0.37$\pm$0.03  \\
   \multirow{1}*{Charm} &6.33$\pm$2.06   &  0.53$\pm$0.19   & 0.36$\pm$0.09  \\
    \hline
  \end{tabular}

  \label{table:parameters}
\end{center}
\vspace{-10pt}
\end{table}

\section{Summary and discussions}
In this work, based on the leading power NRQCD factorization formalism and the LBT model, we have presented a novel finding, namely: the high-$p_\text{T}$ $J/\psi$ production in AA collisions provides a robust probe to the gluon jet quenching. This finding is supported by the following two key results: (1) The agreement between the GFIP event generator and the CMS and ATLAS data for pp collisions, where the gluon fragmentation dominates ($>$ 85$\%$) the high-$p_\text{T}$ $J/\psi$ production over a wide range of $p_\text{T}$; (2) The agreement on the nuclear modification factor $R_{\text{AA}}$ and $v_2$ between the LBT simulation and CMS and ATLAS data for AA collisions, where we find that the suppression of high-$p_\text{T}$ $J/\psi$ production is mainly driven by the gluon energy loss effect. To further confirm our finding and to extract quantitatively the parton energy loss distribution, we have used the data-driven Bayesian analysis for the experimental data on high-$p_\text{T}$ $J/\psi$ suppression. In consistency with the LBT simulation, the extracted $R_{\text{AA}}$ is dominated by the gluon energy loss. We have quantitatively extracted, for the first time, the fraction of average energy loss $\langle \Delta p_\text{T}\rangle/ p_\text{T}$ as well as  the energy loss distributions $W(x)$ for the gluons in QGP. These results help explain the latest high-$p_\text{T}$ $J/\psi$ data, advance the efforts to separately extract gluon energy loss, and mark an important step toward probing the fundamental color structures of QGP medium. One interesting future direction is to combine high-$p_\text{T}$ $J/\psi$ suppression with open heavy flavor and light hadron suppressions to quantify quark versus gluon  energy loss distributions  and test the parton-type dependence of jet-medium interactions~\cite{Arleo:2017ntr}.

{\bf Acknowledgments}

We thank W. Chen for providing 3+1D hydro profile of the bulk medium in our LBT simulation. This research is supported by National Natural Science Foundation of China (NSFC) under Grants No. 12035007, No. 12022512, No. 12147131, No. 12225503, No. 11890710, No. 11890711, No. 11935007, by Guangdong Major Project of Basic and Applied Basic Research No. 2020B0301030008, S.Z. is supported by the MOE Key Laboratory of Quark and Lepton Physics (CCNU) under Project No. QLPL2021P01. J.L. is supported by NSF Grant No. PHY-2209183.

{\bf Author contributions}

Hongxi Xing perceived the main idea and designed the overall project. Shan-Liang Zhang performed the calculations and simulations. All of the authors participated in the scientific discussion and interpretation of the research study, and the writing of the manuscript.

\vspace*{.6cm}


\end{document}